# A Custom-Built Ambient Scribe Reduces Cognitive Load and Documentation Burden for Telehealth Clinicians

Justin Morse[1], Kurt Gilbert[1], Kyle Shin[1], Rick Cooke[1], Peyton Rose[1], Jack Sullivan[1], Angelo Sisante[1]

[1] Included Health, San Francisco, CA 94111

## Abstract

Clinician burnout has motivated the growing adoption of ambient medical scribes in the clinic. In this work, we introduce a custom-built ambient scribe application integrated into the EHR system at Included Health, a personalized all-in-one healthcare company offering telehealth services. The application uses Whisper for transcription and a modular in-context learning pipeline with GPT-4o to automatically generate SOAP notes and patient instructions. Testing on mock visit data shows that the notes generated by the application exceed the quality of expert-written notes as determined by an LLM-as-a-judge. The application has been widely adopted by the clinical practice, with over 540 clinicians at Included Health using the application at least once. 94% (n = 63) of surveyed clinicians report reduced cognitive load during visits and 97% (n = 66) report less documentation burden when using the application. Additionally, we show that post-processing notes with a fine-tuned BART model improves conciseness. These findings highlight the potential for AI systems to ease administrative burdens and support clinicians in delivering efficient, high-quality care.

## 1. Introduction

Clinicians employ the SOAP note format to create comprehensive and standardized summaries of medical visits (Podder et al., 2025). The acronym SOAP represents four essential components of a well-structured medical note: Subjective observations, Objective findings, Assessment, and Plan for follow-up care (Podder et al., 2025). While this format helps produce high-quality documentation, the process of generating detailed SOAP notes for every patient visit can be both time-consuming and challenging (Sinsky et al., 2016). The complexity of modern electronic health record (EHR) systems exacerbates this burden, as navigating these platforms requires significant effort due to their intricate interfaces, frequent updates, and administrative demands (Arndt et al., 2017; Johnson et al., 2021). Consequently, documentation burden has emerged as a leading contributor to clinician burnout (Kumar and Mezoff, 2020; Berg, 2024).

To address the growing issue of burnout and its negative impact on care quality, medical practices have recently turned to AI-powered scribing solutions to streamline the process of medical documentation. As of early 2025, at least 60 AI vendors operate in the ambient medical scribing space, all offering applications that capture visit audio and automatically synthesize clinical documentation for EHR systems (PHTI, 2025). While the long-term impact of AI medical scribes in clinical settings remains uncertain, early findings highlight their potential to reduce documentation time, reduce cognitive load during patient visits, and enhance operational efficiency (Duggan et al., 2025; Tierney et al., 2024). These results have generated excitement about the role of AI in transforming clinical workflows and improving the overall healthcare experience for patients.

While the technology powering AI scribe vendors is typically proprietary, prior research in automated SOAP note summarization leveraged multiple advancements in natural language processing for both audio transcription and note generation. Whisper, an open-source transcription model first described in 2022, uses a transformer architecture to produce token-level utterances from raw spectrogram data (Radford et al., 2023). Since publication, Whisper has been used across multiple domains for batch and realtime transcription applications (Le et al., 2023; Machacek et al., 2023). However, as with other forms of generative AI, hallucinations highlight the need for fine-tuning or prompt engineering to improve performance on domain-specific tasks (Koenecke et al., 2024; Peng et al., 2023).

Early research on SOAP note automation used LSTM-based models to extract key utterances from visit transcripts and generate summaries (Krishna et al., 2021). Subsequent work explored fine-tuned transformers (e.g., BART, T5), with some groups adopting an "extract-and-abstract" approach and others generating notes directly from visit transcripts (Su et al., 2022; Moramarco et al., 2022). More recently, SOAP notes generated by a modified transformer with a specialized cross-attention layer were shown to be more accurate than those generated by GPT-3.5 (Ramprasad et al., 2023). As of March 2025, few studies have explored using the latest class of commercially available decoder models to produce SOAP notes (Yi et al., 2024).

In this work we present a custom-built AI scribe application that is integrated directly into the EHR for the entire medical practice at Included Health. The application employs a modular chain-of-thought prompting strategy with GPT-4o to generate SOAP notes with minimal latency. We demonstrate that these notes only require minor edits in a production telehealth setting. Finally, we show that surveyed clinicians self-report a positive impact on both cognitive load during visits and overall documentation burden.

## 2. Methods

### 2.1 Datasets

We use a publicly available dataset of mock patient visits, Primock57 (Korfiatis et al., 2022), for the characterization of commercial transcription vendors and our in-context learning (ICL) prompting approach. This dataset, accessible via GitHub, contains data from 57 mock visits covering a variety of common medical topics relevant to telehealth clinical care. Each mock visit includes recordings of provider and patient audio, human-written transcripts, and expert-written SOAP notes. For each mock visit, we combine the split provider and patient audio files into a single webm-formatted file before machine transcription.

To analyze clinician adoption of our ambient scribe application, we use institutional data from Included Health, collected over the period of November 2024 to March 2025. This dataset contains information about when a medical encounter occurred, and if the administering clinician used the scribe application to write medical documentation.

Finally, to fine-tune a sequence-to-sequence transformer model (BART) we use another proprietary dataset from Included Health comprised of 130,000 SOAP notes generated by the scribe application paired with their clinician-edited versions. Before fine-tuning, we preprocess the clinician-edited text to remove whitespace and phrases indicating location or attestations.

### 2.2 Transcription

This study evaluates the performance of multiple commercially available transcription models in the medical domain.

**Whisper**: Whisper is a speech-to-text transformer model that generates text from spectrogram data. In this work, we access the Whisper model via OpenAI's Audio API.

**GPT-4o Transcribe** Launched in March 2025, GPT-4o Transcribe is available in two versions: a default version and a "mini" version. Like Whisper, GPT-4o Transcribe is based on a transformer architecture. According to internal benchmarking by OpenAI, GPT-4o Transcribe demonstrates a lower word error rate (WER) across multiple languages compared to Whisper (OpenAI, 2025). This work accesses the GPT-4o Transcribe models through OpenAI's Audio API.

**Evaluation Methodology** To assess transcription fidelity, we analyze 32 audio recordings from the Primock57 dataset. For each transcription model, we calculate a distribution of WER by comparing the machine-generated transcripts to human-written transcripts, which serve as the ground truth. Sample means are compared for statistical significance using a paired Student's

T-test, with alpha = 0.05. Unless otherwise noted, all results are shown as sample means +/- sample standard deviation.

## 2.3 Prompt Engineering

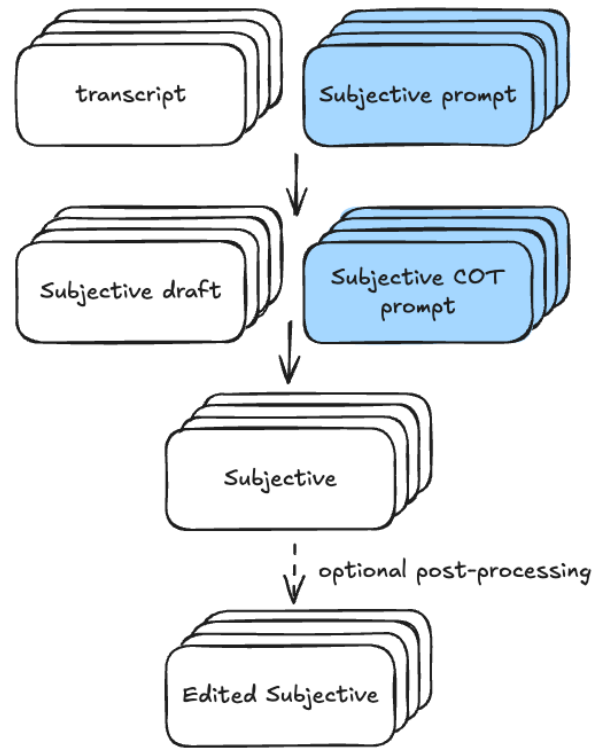

**Figure 1:** A modular chain-of-thought prompting approach to generate SOAP sections in parallel

We use in-context learning with GPT-4o to generate SOAP notes. Whisper visit transcripts are fed with a multi-shot prompt to GPT-4o via OpenAI's Chat Completion API. Instead of relying on a single prompt to create the entire SOAP note, we use a modular approach to generate the following mutually exclusive subsections in parallel (Figure 1 and Appendix A):

1. **Subjective - History of Present Illness:** Captures the patient's chief concern, symptoms, and the context of their visit.
2. **Subjective - Past Medical Encounters and Vitals:** Summarizes biometric data collected during the visit and findings from prior medical visits occurring in the past six months.
3. **Assessment and Plan:** Summarizes the clinician's diagnostic impressions and treatment plan as discussed with the patient.

To improve accuracy and coherence of the note, each generated subsection is subsequently chained to a verification prompt, and again passed to GPT-4o, for a total of six API calls. The subsection from each chain-of-thought reasoning step is stitched together to form the SOAP note.

We similarly generate a long-form Patient Instructions document using multi-shot in-context learning with GPT-4o. The Objective section of the SOAP note, which primarily relies on clinical observations and physical examination, is not generated using AI.

**Evaluation**

To determine if the in-context learning pipeline generates quality SOAP notes, we follow three approaches. First, we generate 32 SOAP notes using the Whisper transcripts of Primock57 visits, and then compare those to the provided expert-written notes using an LLM-as-a-judge. The judging LLM is presented with both the ICL-generated and expert-written notes, along with a rubric (Appendix C) outlining the criteria for scoring the History of Present Illness subsection for each note. The judge is then tasked with selecting the note with the highest overall quality. We then measure win rate, defined as the ratio of instances the note generated by the ICL pipeline was selected divided by the total number of comparisons.

Second, we gather user feedback directly from Included Health clinicians through a survey conducted via Google Forms. The survey includes two questions designed to assess clinician burnout.

Finally, we use automated metrics to evaluate the quality of SOAP notes generated by the scribe application, using the notes submitted to the EHR by clinicians as the ground truth. To determine edit rate, we normalize Levenshtein distance to the character length of the ICL-generated SOAP note. We calculate F1 BERTScore (Zhang et al., 2020) using the bert-score python package with the "bert-base-uncased" model type.

## 2.4 Fine-tuning

To attempt learning the edits most commonly made by clinicians to the History of Present Illness (HPI) subsection generated by the ICL pipeline, we fine-tune BART (Lewis et al., 2020). We use the BART-base model available from the HuggingFace transformer library, with the tokenizer's maximum length set to 512 tokens. We allow each model parameter to be updated during gradient descent.

We trained the model using a proprietary dataset consisting of ~130,000 ICL-generated notes and the corresponding clinician-edited version submitted to the EHR. Training was conducted on an Apple M3 chip, using a batch size of 8 and running for 5 epochs. We employed a cosine learning rate schedule with a warmup ratio of 0.1.

# 3. Results and Discussion

## 3.1 Prompting Whisper improves transcription fidelity of medical audio

To evaluate the accuracy of transformer-based transcription models in the medical domain, we test three commercially available models from OpenAI (Whisper, GPT-4o transcribe, and GPT-4o mini transcribe) using audio recordings from the Primock57 dataset. Of the three models, Whisper achieves the lowest WER ($p < 0.01$, Table 1).

Building on prior studies (Peng et al., 2023), we hypothesized prompting Whisper with key medical terms terms could enhance transcription fidelity of medical audio. To test this hypothesis, we extract the top 200 terms using TF-IDF from an out-of-sample corpus of human-written Primock57 transcripts and incorporate these terms as a prompt for Whisper. The prompted Whisper model achieves a 19% reduction in WER compared to baseline Whisper ($p < 1e-7$, Table 1).

| Model | WER (mean +/- stdev) |
|:---:|:---:|
| Whisper | 0.26 +/- 0.04 |
| Whisper-prompted | **0.21 +/- 0.04** |
| GPT4o-transcribe | 0.31 +/- 0.09 |
| GPT4o-transcribe-mini | 0.29 +/- 0.06 |

**Table 1:** Comparison of WER for various transformer-based transcription models on mock medical audio data

## 3.2 Summarization of medical visits with in-context learning

Next, we investigate the ability of commercially available large language models to produce high-quality SOAP note summaries of medical visits. Leveraging Whisper transcripts of Primock57 visits, we produce 32 SOAP notes using the ICL pipeline outlined in the methods section. When evaluated by rubric-based LLMs-as-a-judge, the SOAP notes generated by the ICL approach achieve a higher win rate compared to expert-written notes (Table 2).

| LLM-Judge | Win rate (+/- stdev) |
|---|---|
| GPT-4o | 0.84 +/- 0.03 |
| Claude 3.7 - sonnet | 0.97 +/- 0.03 |

| LLM-Judge | Win rate (+/- stdev) |
|---|---|
| OpenAI o1-preview | 1 |

**Table 2:** LLM-as-a judge win rate of ICL-generated SOAP notes

Qualitative analysis of the ICL-generated notes (Appendix A) supports the LLM-judges and highlights several appealing features for large medical practices; the notes adhere to a consistent style, they limit the use of abbreviations, and they accurately document critical details of the visit, including pertinent negatives. Additionally, the corresponding patient instructions are written in patient-accessible language while educating and reminding patients about important aspects of their care.

## 3.3 An ambient medical scribe application in production

### System design

A user interface for recording visits and generating notes was integrated into a React-based application already in use by IH clinicians for patient care, ensuring that clinicians could access the new feature without disrupting their existing workflows. On the backend, a GoLang service orchestrates multiple API calls to an internal proxy layer (Figure 2), which routes requests to the appropriate transcription or LLM vendor. The system delivers SOAP notes and patient instructions to users with a p50 latency of 14.4 seconds (Figure 3).

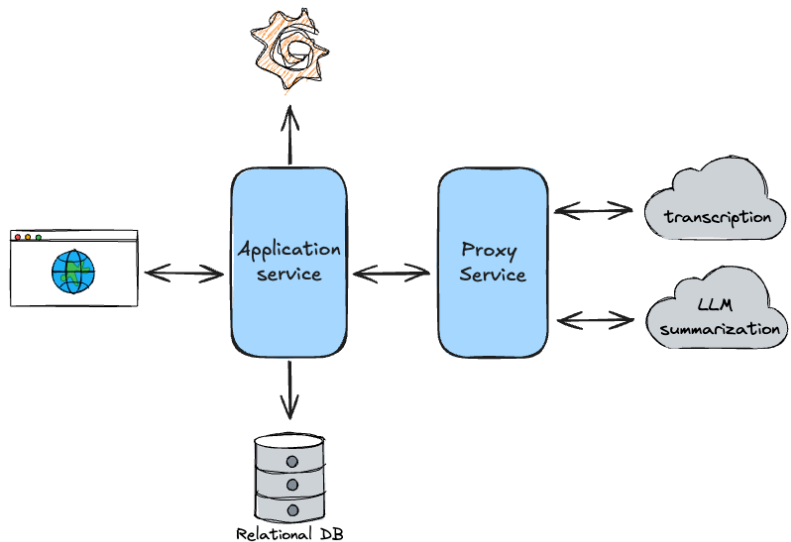

**Figure 2:** System architecture of an ambient medical scribe for use in telehealth applications

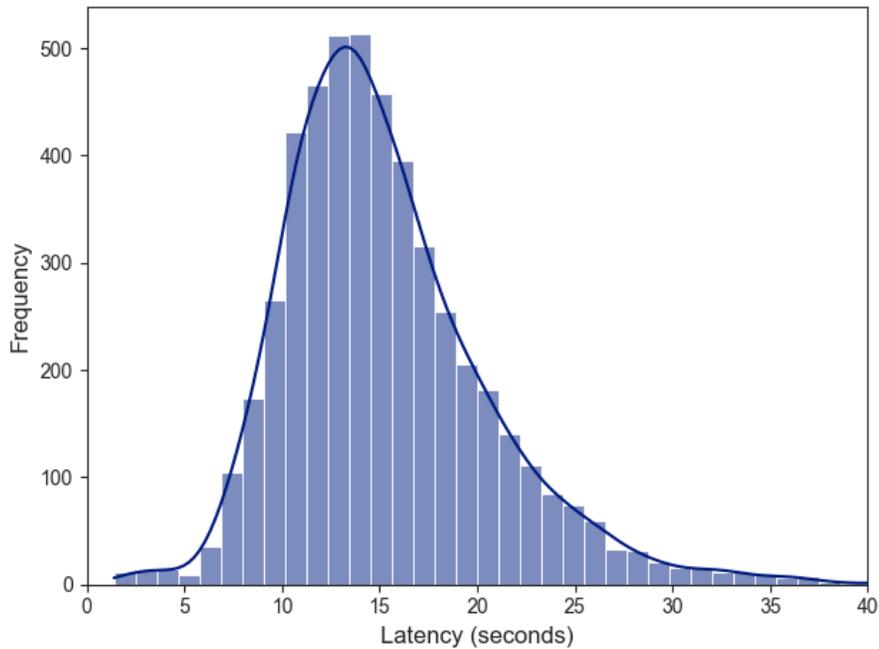

**Figure 3:** Histogram of note generation latency

## User adoption, feedback, and quality at scale

Included Health provides healthcare services across urgent care (UC), virtual primary care (VPC), psychiatry, and therapy. The ambient scribe application was integrated into Included Health's EHR in November 2024. Within three months, the application was used to document nearly 70% of all VPC visits and 40% of UC visits (Figure 4). Adoption in therapy and psychiatry settings has been slower, stabilizing at around 10% two months after launch. To date, over 540 clinicians across all service lines have used the application at least once.

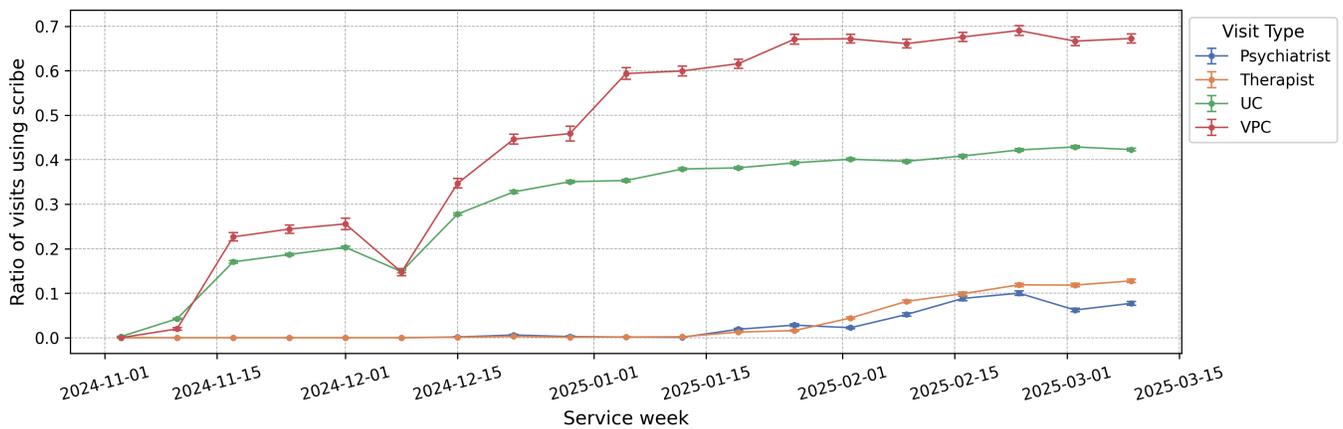

**Figure 4:** Adoption of an ambient medical scribe application across clinical service lines at Included Health

To assess the application's impact on burnout we conducted a survey of all clinicians at Included Health. Among respondents, 94% (n = 63) agreed that the application reduced their cognitive load during patient visits. Additionally, 97% (n = 66) reported that the application decreased their overall documentation burden.

To evaluate the quality of SOAP notes generated by the scribe application in production, we compared the ICL-generated SOAP notes to the final versions submitted by clinicians to the EHR. Since clinicians are required to thoroughly review and edit each scribe-generated note for accuracy, we treat the clinician-submitted notes as the ground truth for benchmarking. Focusing specifically on the HPI section, we found that the character length of ICL-generated notes is, on average, 4% longer than the clinician-submitted versions ($p < 1e-10$, Table 3). Combined with the high F1 BERT score between the two document types (Table 3), these metrics suggest that clinicians primarily edit the ICL-generated notes to improve conciseness while preserving overall semantic meaning.

## 3.4 Fine-tuned models compress SOAP note length with little impact on semantics

In response to these findings, we explore whether fine-tuning a language model can help reduce editing required by clinicians. Since visit transcripts are not retained and thus not available for model training, we frame this task as a post-processing step that directly refines the output of the ICL pipeline rather than operating directly on visit transcripts themselves.

We fine-tune BART-base, a bidirectional sequence-to-sequence transformer, as its cross-attention mechanism is ideal for tasks where input and output are highly similar. Additionally, its relatively small parameter size makes fine-tuning and inference more computationally efficient. We use a dataset consisting of the ICL-generated SOAP notes paired with the clinician-edited version submitted to the EHR. This dataset should allow the model to learn frequently-occurring edits, potentially correcting common misspellings, improving clarity, and removing irrelevant details. As a proof-of-concept, we focus only on the HPI section of the SOAP note.

After processing the HPI section with the fine-tuned BART model, character length decreases by 17% ($p < 1e-10$, Table 3). In comparison, the F1 BERTscore, again calculated using the submitted documents as the reference, drops by only 5% ($p < 1e-10$, Table 3). This suggests that the post-processing step may be effectively removing repetitive information from the notes. Qualitative analysis supports this conclusion, showing that while the post-processed notes remain highly similar to the original ICL-generated versions, they are less repetitive and contain fewer statements about topics less relevant to the chief complaint. However, the post-processing step also eliminates some pertinent negatives, which may impact the completeness of the notes. Future research will investigate whether clinicians prefer these shorter, post-processed notes in a real-world clinical setting.

| HPI document type | Character length | F1 BERTscore |
|---|---|---|
| Clinician-submitted | 1169 +/- 302 | - |
| ICL-generated | 1215 +/- 273 | 0.97 +/- 0.002 |
| Post-processed | 1005 +/- 225 | 0.92 +/- 0.004 |

**Table 3:** Quantification of HPI character length and F1 BERTscore

## 4. Conclusion

We present an in-context learning pipeline designed to produce high-quality SOAP notes and patient instructions using ChatGPT-4o in a production environment. We demonstrate that the accuracy of Whisper transcriptions for medical audio can be enhanced by including a domain-specific prompt. Additionally, we implement a modular chain-of-thought approach for generating SOAP notes and highlight how a post-processing step using a fine-tuned BART model can effectively shorten note length while maintaining semantic integrity. Lastly, our findings emphasize the potential benefits of integrating AI tools into clinical workflows, including a reduction in self-reported clinician documentation burden and cognitive load.

## 5. Limitations and Ethics

**Limitations**: Aside from Primock57 data, the datasets and code described in this work are proprietary and cannot be shared.
**Competing interests**: All authors are employees at Included Health.
**Data policies and ethics**: Included Health holds a Business Associate Agreement (BAA) with OpenAI and other large language model (LLM) vendors to ensure compliance with HIPAA data privacy and security standards. Patient consent is required before AI is used for visit documentation. Clinicians have been coached to always review AI-generated SOAP notes and Patient Instructions, paying special attention to possible hallucinations. All fine-tuning was conducted on proprietary hardware. All examples shared in the appendix of this work were generated using mock data from Primock57.

## 6. Author Contribution

JM conceived of, designed, and executed all experiments described in this paper. JM, KS, RC, PR and JS designed and engineered the ambient scribe application. AS supported application adoption. KG provided extensive clinical guidance. JM wrote the manuscript. All authors reviewed the paper.

# Appendix A. Prompts

| Section | Prompt |
| --- | --- |
| History of Present Illness | CC: In 10 words or less, state the patient's chief reason(s) for the visit.<br><br>HPI:<br>First, extract the information requested below from the transcript.<br>- Onset, duration, and course of symptoms<br>- Severity and character of symptoms<br>- Suspected exposures, including similar symptoms in close contacts<br>- Aggravating and alleviating factors<br>- Associated symptoms<br>- Relevant past episodes of similar symptoms<br>- Treatments/medications that the patient has taken prior to the visit to manage their symptoms.<br>- Effectiveness of treatments/medications taken prior to the visit.<br>- Patient denial of symptoms relevant to their chief complaint using the format 'patient denies <symptoms go here>'<br>- If patient is female and suffering from UTI-like symptoms, the date of their last menses<br>- Any recent Covid exposures, or the result of any at-home Covid tests.<br>- Any other questions to which the patient explicitly answered no or denied.<br>Then synthesize that information into a well-written and comprehensive history of present illness (HPI) for the patient's current medical issue(s).<br><br>Ensure that your note is extremely detailed and thorough. In general, your response should exceed 200 words. Use medical terminology where appropriate. Your tone should be clinical in nature.<br><br>For any topics not discussed or relevant to the patient's condition(s), simply omit them. Don't mention any gaps in information or express uncertainty.<br><br>If the visit is for a standard follow-up or refill request, that should be reflected in the CC and HPI sections. |
| Past Medical Encounters and Vitals | Recent medical encounters:<br>- Summarize any prior encounters with a health care professional occurring within the past 6 months, not including the current visit.<br>- Leave the section blank if no prior encounters were discussed. |

| Section | Prompt |
|---|---|
| | Vitals:<br>- Record any vital signs collected recently or during the visit. Indicate if the patient had difficulty collecting any vital signs. Leave the section and header blank if no vitals were discussed. |
| Assessment and Plan | Assessment:<br>In 2-3 words, use the content in the transcript to state each of the diagnoses discussed during the visit. Then briefly summarize the provider's rationale for each diagnosis. It is very important to only include diagnoses that were explicitly mentioned during the visit. If no diagnoses were provided, your response should only be: "Patient was not provided a diagnosis".<br><br>If a patient's primary reason for visiting was to get a refill or routine follow-up, the assessment should indicate which pathology or physiological system is being managed if any.<br><br>Plan: Was a treatment plan discussed between the provider and patient in the transcript? If so summarize any action items. Note any information about the following if present:<br>- Which additional testing will be needed.<br>- Outside consultations or referrals to different practitioners.<br>- Required medications or procedures.<br>- If medication is prescribed, state the exact dose if available.<br>- Suggestions for follow-up care, whether virtually or in person.<br><br>If this information is not present, do not include it in the note. If a treatment plan was not provided your response should only be: "Patient was not provided a plan"<br><br>It is absolutely critical that you only include plan items that are explicitly and verbatim stated in the transcript. Do not infer, guess, fabricate or generate any plan items or information that is not directly contained in the transcript.<br>The correct response format is as follows:<br>\<diagnosis 1 > - \<supporting criteria for diagnosis 1><br>\<plan for diagnosis 1><br><br>\<diagnosis 2 > - \<supporting criteria for diagnosis 2><br>\<plan for diagnosis 2><br>etc, for all diagnoses |
| Patient Instructions | Use the attached transcript to generate instructions for the patient. Follow a similar style and format as the provided examples.<br><br>The instructions should: |

| Section | Prompt |
|---|---|
| | - Summarize key topics of discussion from the visit
- Summarize follow up items. Pay attention to details like treatment plans or follow-up visits
- Use a conversational - yet professional - voice
- Limit each paragraph to a single message.
- Make sure to avoid repetition
- Use section titles to alert readers to what is coming.
- Minimize information unrelated to the central concepts in the text.
- Use affirmative sentences most of the time.
- Use negative sentences to emphasize avoiding an action.
- Place the most important information first or last.
- When discussing medical concepts, make sure to use simple language.
- Verify that the spelling of medications and medical terms is correct: common medications for weight loss management are Wegovy, Zepbound and Mounjaro |
| Preamble (included in each prompt except the Patient Instructions) | Attached is a transcript of a medical visit occurring between a clinician and their patient. Use the transcript to provide the requested content below. You have an IQ of 120. Think step by step and be concise. Always adhere to the provided requirements. Use a similar style as the attached examples when formulating your response. Strictly adhere to the information provided in the transcript. |
| Requirements (included in each prompt except for Patient Instructions) | - Do not include any other sections of a SOAP note. Only provide what is requested.
- Only include content found in the attached transcript. Do not make up or infer any information not included in the transcript.
- Do not infer any judgements on the patient's symptoms or conditions.
- Write from the first-person perspective of the clinician
- Always respond in english.
- Do not use the term 'transcript' in your response. Visit is an appropriate substitution.
- Do not use any gendered pronouns like he or she to refer to the patient, using instead 'the patient' or 'they'.
- Avoid repetition.
- Do not include any dialog/commentary before or after the requested information
- Do not include content from the example section in your response
- Verify that the spelling of medications and medical terms is correct: common medications for weight loss management are Wegovy, Zepbound and Mounjaro
- Only use normal text. Do not provide markdown formatting. |
| Verification Prompt | While making as few changes as possible, edit your response according to the below guidelines: |

| Section | Prompt |
|---|---|
| | - Remove statements indicating that a topic "was not mentioned" or "was not discussed".
- Ensure that all pertinent negatives stated in the note can be directly supported by content in the transcript.
- If multiple concerns are discussed, break them out into separate paragraphs.
- In the case where no edits are required, return the original note.
- Do not include any commentary on your edits. |

Each prompt excluding the Verification Prompt contain multi-shot examples of target output sequences. They are omitted here for brevity.

| Section | Prompt |
|---|---|
| | |

# Appendix B. Examples

| Primock57 Identifier | Generated Note | Expert Note (from Primock57) |
|---|---|---|
| day1_consulation01 | **SUBJECTIVE**<br>CC: Diarrhea and abdominal cramping<br><br>HPI: The patient presents with a three-day history of diarrhea, reporting watery and loose stool occurring approximately 6-7 times per day. They describe accompanying symptoms of lower left abdominal cramping characterized as a muscular-like pain that is intermittent. The patient reports a sensation of weakness and shakiness without an actual fever but a subjective feeling of being hot approximately three to four days ago. They experienced vomiting at the onset of symptoms, which has since resolved, with no occurrences of blood in the vomit. The patient denies the presence of blood in stool. A reduction in appetite is noted, though fluid intake is maintained with consumption of soups, smoothies, and other liquid foods. The patient suspects a recent Chinese takeaway meal, consumed four days ago, may have initiated their illness. While no other family members have experienced diarrhea, one of their children presented with vomiting. The patient denies any smoking or alcohol consumption and indicates no history of other gastrointestinal issues apart from well-controlled asthma for which they use an inhaler. The symptoms have required the patient to remain close to the toilet, complicating their ability to work as an accountant over the past three days.<br><br>**ASSESSMENT AND PLAN**<br>Gastroenteritis - supported by the patient's symptoms of diarrhea, | Presenting complaint: I've been having really bad diarrhea for the last 3 days<br><br>Note: 3/7 hx of diarrhea, mainly watery. No blood in stool. Opening bowels x6/day. Associated LLQ pain - crampy, intermittent, nil radiation. Also vomiting - mainly bilous. No blood in vomit. Fever on first day, nil since. Has been feeling lethargic and weak since.<br><br>Takeaway 4/7 ago - Chinese restaurant. Wife and children also unwell with vomiting, but no diarrhea. No other unwell contacts.<br>PMH: Asthma<br>DH: Inhalers<br>SH: works as an accountant. Lives with wife and children. Affecting his ADLs as has to be near toilet often.<br>Nil smoking/etOH<br><br>Imp: gastroenteritis<br>Plan:<br>Conservative management - rest, push fluids, paracetamol if feverish. Recommend |

| Primock57 Identifier | Generated Note | Expert Note (from Primock57) |
|---|---|---|
| | abdominal pain, nausea, and recent consumption of potentially unsafe food.<br>- Advised the patient to maintain hydration, possibly including the use of oral rehydration solutions like diarrohide.<br>- Suggested taking paracetamol, two tablets up to four times a day, for fever and general malaise.<br>- Recommended taking 2-3 days off work to allow for recovery.<br>- Advised follow-up in 3-4 days if symptoms do not improve to consider further diagnostic testing, potentially including stool sample analysis.<br><br>**INSTRUCTIONS**<br>Today, we talked about your recent symptoms, including three days of diarrhea. You experienced increased trips to the bathroom with watery stools, left lower abdominal pain described as cramp-like, and early symptoms of vomiting. You mentioned feeling weak but have not had a fever recently. Additionally, you've had a decreased appetite but can manage liquids. We considered your recent meal at a Chinese restaurant as a possible trigger and diagnosed you with gastroenteritis, a common tummy bug.<br><br>Managing Gastroenteritis:<br>The primary focus is to ensure you stay hydrated. Keep drinking plenty of fluids—at least eight glasses a day. Consider picking up an oral rehydration solution, like Dioralyte, from a pharmacy to help replenish essential minerals and electrolytes. If you're feeling weak or feverish, take paracetamol, two tablets up to four times a day, as needed for relief. It's important to avoid solid foods until your symptoms improve, and gradually reintroduce bland foods like rice, bananas, or toast. | OTC diarolyte. To review in 3-5d if symptoms not improving. To see earlier if feeling more unwell. |

| Primock57 Identifier | Generated Note | Expert Note (from Primock57) |
|---|---|---|
| | Time Off for Recovery:<br>Rest is crucial for your recovery, so I recommend taking the next two to three days off from work. This downtime will allow your body to clear the infection thoroughly. Be sure to avoid activities that may exhaust you or exacerbate your symptoms, and take it easy during this period.<br><br>Follow-Up Actions:<br>Monitor your condition over the next few days. If your symptoms do not improve in three to four days, or if they get worse, please reach out to schedule another visit. We will determine if further testing, such as a stool sample, is necessary to investigate other potential causes. Additionally, if you notice blood in your stool or vomit or experience severe pain, seek medical help promptly. Thank you, and I wish you a speedy recovery. | |
| day1_consulation02 | **SUBJECTIVE**<br>CC: Itchy, sore skin rash<br><br>HPI: The patient presents with a four-day history of itchy and sore skin primarily involving the chest, hands, and interior elbows. The patient describes the rash as very itchy, interfering with both daily work responsibilities and sleep, and identifies it as more severe compared to previous eczema episodes. Previous episodes were limited to the arms. The patient has attempted self-treatment with over-the-counter steroid cream and antihistamines, both of which provided no relief. The patient denies any bleeding, discharge, or pus from the rash but notes that the skin appears cracked and dry in some areas. There are no known new environmental exposures, changes in personal care products, or potential contact with irritants | Presenting complaint: I have sore, red skin<br><br>Note: 4/7 hx of dry itchy skin, mainly on chest and back. Skin is also sore and cracked. No bleeding or discharge. No spread elsewhere. Not aware of any triggers. Used steroid cream OTC last night with minimal effect. Pt has also tried antihistamines. Systemically well otherwise - no fevers, no vomiting, no SOB/cough. No bowel sx.<br><br>PMH: Eczema, Asthma |

| Primock57 Identifier | Generated Note | Expert Note (from Primock57) |
|---|---|---|
| | or allergens.

The patient denies any other systemic symptoms such as fever, breathing difficulties, or urinary issues and confirms regular bowel functions. There are also no known allergies or use of any regular medications. The patient lives with flatmates and admits to occasional alcohol consumption but no smoking. The patient has a history of eczema.

**ASSESSMENT AND PLAN**
Eczema flare-up - supported by the patient's report of itchy, red, and sore skin primarily on the chest and inside the elbows; history of eczema diagnosis; trial of over-the-counter steroid cream with limited relief; background of eczema symptoms more localized to arms.
- I will prescribe a stronger steroid cream for the patient to use.
- The patient should use emollients for moisturizing the skin during bath and shower.
- Advised the patient to use antihistamines such as loratadine or pyriton, and I have prescribed fexofenadine for stronger relief.
- Recommend keeping a diary to identify any potential triggers.
- Suggested follow-up visit in 7 to 10 days if symptoms do not improve for reassessment.

**INSTRUCTIONS**
During our visit, we addressed your recent flare-up of itchy and sore skin, primarily affecting your chest, inside your elbows, and hands. You mentioned difficulty in daily activities due to persistent itchiness, which is similar to but more widespread than your previous eczema episodes. | DHx: Nil regular NKDA
SH: lives with flatmates, works as a pharmaceutical manager. Non-smoker, social EtOH.
ICE: pt thinks might be related to eczema. No concerns

Imp: flare up of eczema
Plan:
1. Steroid cream - Betnovate BD
2. Emollients - cetraben - use up to QDS
3. Trial antihistamines
4. Keep trigger diary
5. Review in 10-14d if symptoms no better, or earlier if any concerns/worsening sx |

| Primock57 Identifier | Generated Note | Expert Note (from Primock57) |
|---|---|---|
| | Eczema Management:
We suspect that this could be a flare-up of your eczema. To manage the symptoms, I am prescribing a stronger steroid cream and emollients to use in the bath and shower. Please use these over the next seven to ten days to help soothe and moisturize your skin. Additionally, it may be helpful to try a stronger antihistamine like fexofenadine, which might give you more relief than previous antihistamines you have tried.

Trigger Identification:
It's important to monitor any potential triggers that might be causing or worsening your symptoms. Consider writing down any changes in your routine, such as new skincare products or different clothing materials, that could contribute to your condition. We'll use this information to discuss further treatment options if necessary.

Follow-Up Actions:
- Start using the prescribed steroid cream and emollients regularly for the next 7-10 days.
- Try fexofenadine for itchiness relief.
- Keep a diary of potential triggers to review in our next appointment.
- Schedule a follow-up consultation in one week to evaluate your symptoms.
- Reach out to us sooner if your symptoms do not improve or if you experience any new issues. | |

# Appendix C. LLM-as-a-judge scoring rubric

Rubric for Scoring a History of Present Illness (HPI)
Chief Complaint:
2 - An excellent HPI clearly identifies the chief complaint, providing a concise and accurate summary of the primary reason for the visit.
1 - A proficient HPI identifies the chief complaint but may lack clarity. If the chief complaint is vague or incomplete, the HPI needs improvement.
0 - An unsatisfactory HPI omits the chief complaint entirely. Writing the CC in language in the words of the patient should not penalized.

Chronology:
2 - A well-written HPI provides a clear and logical timeline of events leading to the current issue, ensuring that the sequence of symptoms and events is easy to follow.
1 - A proficient HPI includes a general timeline but may lack some clarity or detail. If the timeline is unclear or incomplete, with significant gaps in the sequence, the 0 0 - HPI needs improvement. An unsatisfactory HPI either fails to provide a timeline or presents events in a disorganized and confusing manner.

Symptom Description:
2 - An excellent HPI thoroughly describes the patient's symptoms using all relevant dimensions, such as location, quality, severity, duration, timing, context, modifying factors, and associated symptoms.
1 - A proficient HPI describes symptoms but may omit one or two key dimensions or lack some detail. If the symptom description is superficial and omits multiple key dimensions, the HPI needs improvement.
0 - An unsatisfactory HPI provides little to no description of symptoms or misses critical details entirely.

Pertinent Positives and Negatives:
2 - A strong HPI includes all relevant pertinent positives and negatives, which are essential for guiding the differential diagnosis.
1 - A proficient HPI includes some pertinent positives and negatives but may miss a few key details. If the HPI includes only a few pertinent positives and negatives or lacks relevance, it needs improvement.
0 - An unsatisfactory HPI either omits pertinent positives and negatives entirely or includes irrelevant information.

Clarity and Organization:
2 - An excellent HPI is well-organized, concise, and easy to follow, presenting information in a logical and structured manner.

1 - A proficient HPI is mostly organized but may include minor redundancies or unclear phrasing. If the HPI is somewhat disorganized or difficult to follow in places, it needs improvement.
0 - An unsatisfactory HPI is disorganized, unclear, or excessively verbose, making it difficult to understand.

Relevance:
2 - A high-quality HPI includes only information that is directly relevant to the chief complaint and the patient's current condition.
1 - A proficient HPI includes mostly relevant information but may include minor extraneous details. If the HPI includes some irrelevant information or omits important relevant details, it needs improvement.
0 - An unsatisfactory HPI includes mostly irrelevant information or omits critical details entirely.

Contextual Information:
2 - An excellent HPI provides appropriate context, including relevant past medical history, social history, and family history, as they relate to the current issue.
1 - A proficient HPI provides some context but may omit minor relevant details. If the HPI provides limited context or omits significant relevant details, it needs improvement.
0 - An unsatisfactory HPI either fails to provide context or includes irrelevant contextual information.

Professionalism:
2 - A strong HPI uses professional, patient-centered language and avoids judgmental or biased phrasing.
1 - A proficient HPI generally uses professional language but may include minor lapses in tone or phrasing. If the language is occasionally unprofessional or lacks a patient-centered focus, the HPI needs improvement.
0 - An unsatisfactory HPI uses unprofessional, judgmental, or inappropriate language.